\documentclass[12pt]{article}

\usepackage[english]{babel}
\usepackage{fancyhdr}
\usepackage{amsmath}
\usepackage{amssymb}
\usepackage{amsfonts}
\usepackage{psfrag}
\usepackage[applemac]{inputenc}
\usepackage{graphicx,wrapfig}
\usepackage[bf,footnotesize]{caption2}
\usepackage{pstricks}
\usepackage{enumerate}

\newcommand{\beq}{\begin{equation}}
\newcommand{\eeq}{\end{equation}}
\newcommand{\bea}{\begin{eqnarray}}
\newcommand{\eea}{\end{eqnarray}}
\newcommand{\bag}{\begin{align}}
\newcommand{\eag}{\end{align}}

\newcommand{\ie}{${\it i.e.}$}

\newcommand{\TeV}{\,\mathrm{TeV}}

\newcommand{\eq}[1]{Eq.~(\ref{#1})}
\newcommand{\fig}[1]{Fig.~\ref{#1}}
\newcommand{\eqs}[1]{Eqs.~(\ref{#1})}

\newcommand{\Lag}{\mathcal{L}}
\newcommand{\Op}{\mathcal{O}}

\begin{document}
\baselineskip=18pt
\setcounter{footnote}{0}
\setcounter{figure}{0}
\setcounter{table}{0}






\begin{titlepage}
\begin{flushright}
February 2012
\end{flushright}
\vspace{.3in}

\begin{center}
\vspace{1cm}

{\Large \bf  
Probing the SM with Dijets at the LHC}

\vspace{1.2cm}

{\large Oriol Dom\`enech\,$^a$, Alex Pomarol\,$^a$ and Javi Serra\,$^b$\\}
\vspace{.8cm}
{\it {$^a$\,Departament  de F\'isica, Universitat Aut{\`o}noma de Barcelona, 08193~Bellaterra,~Barcelona}}\\
{\it {$^b$\,Department of Physics, LEPP, Cornell University, Ithaca NY 14853}}

\vspace{.4cm}


\end{center}
\vspace{.8cm}

\begin{abstract}
\medskip
\noindent
The LHC has started to explore  the  TeV energy regime, 
probing  the SM  beyond  LEP and Tevatron.
We show how the dijet  measurements at the LHC  are able to  test
certain   sectors of the SM  at an unprecedented  level. 
We provide  the best bounds 
on all possible four-quark interactions
and translate them into limits on the compositeness scale
of the quarks and gluons.
We also provide constraints  on  extra gauge bosons,
$Z'$, $W'$ and $G'$,
 and  on  new interactions proposed 
to  explain the present  measurement of the  forward-backward asymmetry of the top.
\end{abstract}

\bigskip

\end{titlepage}


\section{Introduction}

The LHC is currently exploring  the TeV-energy frontier, 
searching for new states  beyond the Standard Model (SM)  \cite{lhc}.
Although not a precision machine, the LHC 
 is also  sensitive to  deformations of the SM that grow with energy, such as contact-interactions,
  allowing  us then   to  test sectors of the SM  at an unprecedented level.

Among the SM sectors that can be better tested at the LHC than in previous experiments
is the  quark sector.
As we will see, the right-handed quark sector  has been poorly   tested
at LEP and Tevatron,  while its left-handed counterpart  only 
 electroweak symmetry breaking effects have been scrutinized.  
There is definitely room for improvement at the LHC.

The purpose of this paper is to use 
the  dijet angular distributions  at the LHC
\cite{Collaboration:2010eza,  Aad:2011aj,Khachatryan:2010te,Khachatryan:2011as} 
to  improve our understanding of the quark sector of the SM.
By using the observable
$F_\chi$, we will put the best bounds on all   possible four-quark interactions.
These new contact-interactions are present in scenarios in which  the quarks are composite states
\cite{Eichten:1983hw}, a natural possibility that can 
arise in  models where   strong dynamics   are expected to be around the TeV-scale, such as 
 composite Higgs models \cite{Agashe:2004rs,Giudice:2007fh}
or scenarios with low-energy supersymmetry breaking \cite{Gabella:2007cp}.
Dijets at the LHC provide then  the best bounds on some of these scenarios. 
We will also show how our dijet analysis constrains
new heavy gauge sectors beyond the SM,
and the degree of compositeness of the SM gauge bosons, most significantly
that of the gluon.

The article is organized as follows.
We will start showing how well the different sectors of the SM  have been tested  in the pre-LHC era,
and which  type of deformations from the SM, parametrized by dimension-six operators,  
can be  much better bounded by the LHC.
In section~\ref{dijets} we present the calculation of the dijet angular distributions
and the $F_\chi$ parameter that seems to be very well suited to  obtain limits
on four-quark interactions.
By using the LHC data we will  obtain bounds on all possible
new four-quark interactions (see also  Ref.~\cite{Bazzocchi:2011in}), and translate these bounds
 into  limits  on the scale of compositeness of the  quarks,
on possible new gauge sectors,  on the 
scale of compositeness of the  SM gauge bosons, and finally 
on new physics scenarios 
responsible for explaining the present experimental discrepancy in the forward-backward asymmetry $A_{FB}$ of the top
 \cite{Aaltonen:2011kc,Abazov:2011rq}.

\section{Tests of the SM sectors  before  LHC}  
\label{sec1}

Let us consider a new sector beyond the SM (BSM)
whose physical scale, generically referred by $\Lambda$ (that, for example, can be associated with  the  mass of the new states), is  assumed to be much larger than the momenta  $p$  at which we are probing the SM, $\Lambda\gg p$.
We can then parametrize  the deviations from the   SM by
   higher dimensional operators added to the SM Lagrangian \cite{Buchmuller:1985jz}:
\beq
\Lag_{eff} = \Lag_{SM} + \frac{c_i}{\Lambda^2} \Op_i^{d=6} + \cdots\, ,
\label{dim6}
\eeq
where we  only keep the dominant contributions  corresponding to operators of dimension six,
 assuming lepton and baryon number conservation.
Among these operators it is important to distinguish between two classes:
\begin{enumerate}
\item Operators involving extra powers of SM fields:
\beq
(\bar q_L\gamma_\mu q_L)^2
\ , \ \ 
(\bar{q}_L \gamma_\mu q_L)(H^\dagger D^\mu H) 
\ ,\   ...
\label{1st}
\eeq
\item Operators involving extra (covariant) derivatives:
\beq
\label{2nd}
\bar q_L\gamma_\mu q_L D_\nu F^{\mu\nu}
\ , \ \ 
\bar q_L u_R D_\mu D^\mu H
\ ,\    ...
\eeq
\end{enumerate}
The coefficients $c_i$ in front of the first class of operators 
are parametrically  proportional to the square of  a  coupling of the SM fields
to the BSM sector,  
and then they  can be as large as    $c_i\lesssim 16\pi^2$.
On the other hand,  the coefficients $c_i$ of the second class of  operators
should not contain  couplings  and are expected to be of order one  $c_i\lesssim O(1)$.
 This distinction is important  when considering strongly-coupled BSM
 with part of the   SM fields arising as composite states of this new sector. 
In this case $\Lambda$  corresponds to the mass of the   heavy resonances  of the new strong sector whose
 couplings, referred  as $g_\rho$, can be as large as  $ \sim 4 \pi$.
Hence operators of the first class with $c_i\sim g^2_\rho$ give  generically  more 
significant modifications to  SM physics than those of the second class 
\cite{Eichten:1983hw,Giudice:2007fh,topphysics}.

At present we have important constraints on  $c_i/\Lambda^2$ 
coming from precision measurements of  SM  observables.
Let us start considering  those involving  SM  fermions.
In the Appendix we give the full list of independent operators involving quarks.
Neglecting fermion masses (chiral  limit),  we have
  that  the impact of the dimension-six operators
  on  SM physical processes  can  generically 
  be parametrized   by two new  types of  interactions:
  \begin{equation}
\frac{\alpha_\psi}{\Lambda^2}(\bar\psi\gamma_\mu\psi)^2
+ \beta_\psi \frac{v^2}{\Lambda^2}  A_\mu\bar\psi\gamma^\mu\psi\, \, ,
\label{dev}
\end{equation}
where we denote  collectively by  $A_\mu=W_\mu,Z_\mu,...$ the SM gauge bosons, by $\psi=u_L,u_R,...$ the SM fermions of a given chirality, 
 $v\simeq 246$ GeV  is the Higgs vacuum expectation value (VEV),
and  $\alpha_\psi$ and $\beta_\psi$ measure the strength of the interactions.
Since  both types of interactions in  \eq{dev} can arise from operators of the first class (\eq{1st}),
one has $0< |\alpha_\psi|,|\beta_\psi|\lesssim 16\pi^2$.
The first term of   \eq{dev}  gives contributions to 
   four-fermion processes that scale as $p^2/\Lambda^2$ 
  where $p$ characterizes the momenta of the process.
The second term   corresponds to 
 deviations from the SM gauge interactions at zero-momentum,
and therefore these can only  arise for the $W$ and  $Z$ gauge boson and must be proportional to the 
Higgs VEV.

In principle, very stringent  constraints on new interactions of the type of  \eq{dev} 
arise from flavor physics \cite{Bona:2007vi}. 
It is not our purpose here to discuss them;  
they are  very model dependent
and can be avoided if  a flavor symmetry is imposed  in the BSM sector. 
For example
we can assume a flavor symmetry for the three left-handed quarks $q_L$,  the three right-handed down-quarks $d_R$, and the two lightest right-handed up-quarks $u_R$, given by 
\beq
G_F\equiv U(3)_{q} \otimes U(3)_{d} \otimes  U(2)_{u}\, ,
\label{gf}
\eeq
 and similarly for the lepton sector. 
 Due to  the absence of important constraints on the flavor physics for the  right-handed  top $t_R$, we can consider it a singlet of the flavor symmetry.
This allows us to treat the $t_R$ independently of the other quarks; 
 its  physical implications, some of them already studied in Ref.~\cite{topphysics},
are left   for  a future publication.
Yukawa couplings   break the $G_F$ symmetry, but it can be shown, by using
a spurion’s power counting,
that flavor constraints on
dimension-six operators can be easily satisfied for  $\Lambda$ slightly above the electroweak scale \cite{Bona:2007vi}. From now on, we will consider BSM that, up to Yukawa couplings, fulfill
the flavor symmetry $G_F$.

At LEP the properties of the leptons $\psi=l_L,l_R$   were very well measured, 
putting bounds at the per-mille level on deviations from the SM predictions
either arising from vertex corrections or  new four-lepton contact interactions.
From  \cite{Nakamura:2010zzi}, one gets  $\Lambda/(\sqrt{|\alpha_{l_{L,R}}|}), \Lambda/(\sqrt{|\beta_{l_{L,R}}|}) \gtrsim 3-4$   TeV.
This implies, for example,  
that the scale of compositeness of the leptons is larger than  $40-50$ TeV
for  $\alpha_{l_{L,R}}\sim g_\rho^2\sim 16\pi^2$.
Thus, the leptonic sector  
has been very well tested at LEP and 
recent LHC  data, having only quarks in the initial state, cannot 
provide better bounds.

For the left-handed quark sector $\psi=q_L$,  
there  are  very  strong  constraints on interactions of the second type of \eq{dev}.
The most important ones arise from  Kaon and $\beta$-decays
\cite{Nakamura:2010zzi} which have allowed to measure very precisely quark-lepton universality of the $W$ interactions.
This leads to
bounds on deviations from the $W$ coupling to left-handed quarks  as strong as those for leptons, 
$\Lambda/(\sqrt{|\beta_{q_L}}|) \gtrsim 3-4$  TeV,  which we  do not expect  to be improved substantially at the LHC. 
Similar limits are obtained from measurements at LEP of the $Z$ decay to hadrons \cite{Nakamura:2010zzi}.
Bounds on four-$q_L$ interactions are weaker, with the main constraint coming from Tevatron and giving
$\Lambda/(\sqrt{|\alpha_{q_L}|}) \gtrsim 1  \TeV$ \cite{Nakamura:2010zzi}.  
Clearly, the LHC can increase these bounds 
considerably  as we will show later.
While theories
of composite Higgs and composite $q_L$, where one expects large $\alpha_{q_L}$ and $\beta_{q_L}$ coefficients
(since $\alpha_{q_L}\sim \beta_{q_L}\sim g_\rho^2\lesssim16\pi^2$) \cite{Eichten:1983hw,Giudice:2007fh,topphysics},
are very constrained by present experimental data,
 theories with only $q_L$ composite (and  elementary Higgs, as those for example in Ref.~\cite{Gabella:2007cp}) where  only $\alpha_{q_L}$ is expected to grow with $g^2_\rho$,
 are  not so constrained.
LHC dijets can then, as we will see, probe these scenarios at an unprecedented level.

Regarding right-handed quarks $u_R$ and $d_R$, 
 their couplings to gauge bosons are still poorly measured, due to their small coupling to $W$ and $Z$. 
For example, one of the best bounds, arising from LEP, are  on the $Z$ coupling to $b_R$ which reads $0 \lesssim \delta g_{b_R}/g_{b_R}\lesssim 0.2$ \cite{Kumar:2010vx}.
 Furthermore these  vertices can be protected by  symmetries of the BSM sector \cite{Agashe:2006at}.
The strongest constraints on $\alpha_{u_R,d_R}$  are again coming from Tevatron 
and,  as for  the left-handed case,  LHC can  improve them significantly.
  Similar conclusions  have been recently reached  in Ref.~\cite{Redi:2011zi}.

For completeness, we  comment on  chirality-flip processes that are sensitive to  fermion masses.
For  $m_\psi=Y_\psi v\not=0$,  two new types of interactions can be  added to \eq{dev}:
 \beq
  \gamma_\psi \frac{m_\psi}{\Lambda^2} F^{\mu\nu}\bar\psi\sigma_{\mu\nu}\psi+\delta_\psi \frac{Y_\psi^2}{\Lambda^2}(\bar \psi\psi)^2\, ,
\label{devm}
\eeq
where $\gamma_\psi$  and $\delta_\psi$   are coefficients of order one.
  Concerning the first one, Re$[\gamma_\psi]$   and Im$[\gamma_\psi]$ give a contribution  
to the  magnetic  and electric dipole moment respectively.
Only electric dipole moments
give important constraints  on BSM sectors,
but they can be avoided 
by demanding CP-invariance  in the BSM. 
Moreover, in  most of the   BSM they arise at the 
one-loop level.
The second interaction in \eq{devm} corresponds to new four-fermion interactions,
but suppressed with respect to those of \eq{dev} by Yukawa couplings,
hence we will not consider them in this work.

Let us also mention that bounds on the interactions \eq{dev} can also constrain
BSM contributions to  the self-energies of the SM gauge bosons.
These effects can be parametrized by  five quantities
$\widehat  S$, $\widehat T$, $W$, $Y$ and $Z$ \cite{Barbieri:2004qk}.
The first two, $\widehat  S$, $\widehat T$, are proportional to $v^2/\Lambda^2$
and find their best bound from LEP and Tevatron data.
 The  $W$ and $Y$ parameters, that measure the compositeness of the $W_\mu$, $Z_\mu$ and photons,
  are also bounded by  LEP data
 at the per-mille level, but since these effects  grow with the momenta as $p^2/\Lambda^2$,
we can expect   LHC to improve the bounds.
Also at the LHC  the best bound on the $Z$-parameter, that measures the compositeness of the gluon,
can be obtained.

We then conclude that  BSM physics generating  four-quark interactions are not
severely constrained by the pre-LHC data.  
Especially interesting BSM scenarios that contribute to this type of  interactions
are theories  of composite quarks, either
composite $u_R$ and $d_R$ (and Higgs),  composite $q_L$ (if the Higgs is elementary), or composite gluons.
Below we will  show how dijets at the LHC  constrain these scenarios.

\section{Dijets at the LHC}
\label{dijets}
The study at the LHC of the angular distributions of dijets in the process $pp \rightarrow jj$ has been shown to be a powerful tool to constrain the size of four-quark contact interactions \cite{Collaboration:2010eza,Aad:2011aj,Khachatryan:2010te,Khachatryan:2011as}. 
Here we will follow these analyses to put constraints on all possible four-quark interactions.
Out of  the complete list of dimension-six operators in \ref{4fer}, only those involving the 
first family of quarks, up and down, are relevant for our analysis.
The reason is the following. 
In $pp \rightarrow jj$  the dominant contributions at high dijet invariant-mass  $m_{jj}$ 
arise from valence-quarks  initial states, \ie, $uu, dd, du$,
 being  $u\bar u$ or $uc$ initial state processes very suppressed. 
 For example, in the SM we have
\beq
\left( \frac{\sigma(u \bar u \rightarrow u \bar u)}{\sigma(u u \rightarrow u u)} \right)_{SM}^{m_{jj} > 2 \TeV} \simeq
0.04 \ , \quad\quad  \left( \frac{\sigma(u c \rightarrow u c)}{\sigma(u u \rightarrow u u)} \right)_{SM}^{m_{jj} > 2 \TeV} \simeq   0.01\, .
\eeq
Furthermore, processes with other families in the final states but
having  $u,d$ in the initial state, such as   $uu\rightarrow ss,cc$,
do not arise from the four-quark operators  of \ref{4fer} due to the flavor symmetry  $G_F$.
We are then led to consider  partonic processes  involving only   first family quarks,
 $uu\rightarrow uu$, $dd\rightarrow dd$ and $ud\rightarrow ud$,
that allow us to  reduce the set of operators of \eq{O8F} to
\bea
\label{tOuu1}
\Op^{(1)}_{uu}&=&(\bar{u}_R\gamma^\mu u_R)(\bar{u}_R\gamma_\mu u_R)\nonumber\\
\Op^{(1)}_{dd}&=&(\bar{d}_R\gamma^\mu d_R)(\bar{d}_R\gamma_\mu d_R)\nonumber\\
\Op^{(1)}_{ud}&=&(\bar{u}_R\gamma^\mu u_R)(\bar{d}_R\gamma_\mu d_R)\nonumber\\
\Op^{(8)}_{ud}&=&(\bar{u}_R\gamma^\mu T^A u_R)(\bar{d}_R\gamma_\mu T^A d_R)\nonumber\\
\Op^{(1)}_{qq}&=&(\bar{q}_L\gamma^\mu q_L)(\bar{q}_L\gamma_\mu q_L)\nonumber\\
\Op^{(8)}_{qq}&=&(\bar{q}_L\gamma^\mu T^A q_L)(\bar{q}_L\gamma_\mu T^A q_L) \nonumber\\
\Op^{(1)}_{qu} &=& (\bar{q}_{L} \gamma^\mu q_{L}) (\bar{u}_{R} \gamma_{\mu}  u_R) \nonumber\\
\Op^{(8)}_{qu} &=& (\bar{q}_{L} \gamma^\mu T^A q_{L}) (\bar{u}_{R} \gamma_{\mu}  T^A u_R) \nonumber\\
\Op^{(1)}_{qd} &=& (\bar{q}_{L} \gamma^\mu q_{L}) (\bar{d}_{R} \gamma_{\mu}  d_R) \nonumber\\
\Op^{(8)}_{qd} &=& (\bar{q}_{L} \gamma^\mu T^A q_{L}) (\bar{d}_{R} \gamma_{\mu}  T^A d_R)
\label{4u}
\eea
where here we do not sum over flavor indices and from now on $q_L=(u_L,d_L)$.
Apart from \eq{4u}, there are  other dimension-six operators  (see the lists of \ref{qv}, \ref{2ndclass} and \ref{YukawaOp}) 
that can  contribute to  dijets. 
Nevertheless, these other operators are  either suppressed 
by $v^2/p^2$  or  Yukawa couplings with respect to  those of \eq{4u},
or  can be rewritten,  by use of  equations of motion, as  four-quark operators plus
other operators not relevant for dijet physics.
Therefore \eq{4u} exhausts the list of all leading operators   contributing to dijets.

At the partonic level   the SM differential cross section of
$pp \rightarrow jj$  is dominated by  QCD interactions \cite{Eichten:1984eu}:
\bea
\frac{\hat s^2}{\pi \alpha_s^2} \frac{d \sigma}{d \hat t} \left( q_iq_i \rightarrow q_iq_i \right)_{SM} &=&
\frac{4}{9} \frac{\hat s^2 + \hat u^2}{\hat t^2} + \frac{4}{9} \frac{\hat s^2 + \hat t^2}{\hat u^2} - \frac{8}{27} \frac{\hat s^2}{\hat t \hat u}\, ,\ \ \ \ \ \  \ \ \ (q_i=u,d)\nonumber \\ 
\frac{\hat s^2}{\pi \alpha_s^2} \frac{d \sigma}{d \hat t} \left( ud \rightarrow ud\right)_{SM} &=& 
\frac{4}{9} \frac{\hat s^2 + \hat u^2}{\hat t^2}\, ,\nonumber \\  
\frac{\hat s^2}{\pi \alpha_s^2} \frac{d \sigma}{d \hat t} \left( gq_i \rightarrow gq_i \right)_{SM} &=& 
(\hat s^2 + \hat u^2) \left( \frac{1}{\hat t^2} - \frac{4}{9} \frac{1}{\hat s \hat u} \right)\, ,\nonumber\\  
\frac{\hat s^2}{\pi \alpha_s^2} \frac{d \sigma}{d \hat t} \left( gg \rightarrow gg \right)_{SM} &=& 
\frac{9}{2} \left( 3 - \frac{\hat t \hat u}{\hat s^2} - \frac{\hat s \hat u}{\hat t^2} - \frac{\hat s \hat t}{\hat u^2} \right)\, ,\nonumber\\
\frac{\hat s^2}{\pi \alpha_s^2} \frac{d \sigma}{d \hat t} \left( gg \rightarrow q_i \bar{q}_i \right)_{SM} &=&
\frac{3}{8} (\hat t^2 + \hat u^2) \left( \frac{4}{9} \frac{1}{\hat t \hat u} -  \frac{1}{\hat s^2} \right)
\, ,
\label{smjj}
\eea
where  $\hat s$, $\hat t$ and $\hat u$ are the partonic  Mandelstam variables, and we are working in
the massless quark limit.
Contributions from the operators of \eq{4u}  give
\bea
\frac{d\sigma}{d \hat t}(q_iq_i\rightarrow q_iq_i)_{BSM} &=&
-\frac{8\alpha_s}{27\Lambda^2}\bigg[A_1^{q_i} \frac{\hat{s}}{\hat{t}\hat{u}}-A_2^{q_i}\bigg(\frac{\hat{u}^2}{ \hat{t}}+\frac{\hat{t}^2}{\hat{u}}\bigg)\frac{1}{\hat{s}^2}\bigg]+\frac{4}{27\pi\Lambda^4}\bigg[B_1^{q_i}+B_2^{q_i}\, \frac{\hat{u}^2+\hat{t}^2 }{\hat{s}^2}\bigg]\, ,\nonumber\\
\frac{d\sigma}{d \hat t}(ud\rightarrow ud)_{BSM} &=&
\frac{ 2\alpha_s}{9\Lambda^2}\bigg[A_3\frac{1}{\hat{t}}+A_4\frac{\hat{u}^2}{\hat{s}^2\hat{t}}\bigg]+\frac{1}{36\pi\Lambda^4}\bigg[ B_3+B_4\,\frac{\hat{u}^2}{\hat{s}^2}\bigg]\, ,
\label{bsmjj}
\eea
where
\bea
\nonumber
A_1^{u,d}&=&\frac12 c^{(8)}_{qq}+\frac32 (c_{uu,dd}^{(1)}+c^{(1)}_{qq})\, , \\
\nonumber
A_2^{u,d}&=&\frac34 c^{(8)}_{qu,qd}\, , \\
\nonumber
A_3&=&\frac12 (2c^{(8)}_{qq}+c^{(8)}_{ud} )\, , \\
\nonumber
A_4&=&\frac12 (c^{(8)}_{qu}+c^{(8)}_{qd})\, , \\
\nonumber
B_1^{u,d}&=&(A_1^{u,d})^2+\frac{1}{16} \bigg[c^{(8)}_{qq}+3(c^{(1)}_{qq}-c^{(1)}_{uu,dd}) \bigg]^2\, , \\
\nonumber
B_2^{u,d}&=&\frac{3}{16} (c^{(8)}_{qu,qd})^2+\frac{27}{32}(c^{(1)}_{qu,qd})^2\, , \\
\nonumber
B_3&=&A_3^2+\frac{1}{4}(2c^{(8)}_{qq}-c^{(8)}_{ud})^2+\frac98(2c^{(1)}_{qq}+c^{(1)}_{ud})^2+\frac{9}{8}(2c^{(1)}_{qq}-c^{(1)}_{ud})^2\, , \\
B_4&=& 
A_4^2+\frac{1}{4}(c^{(8)}_{qu}-c^{(8)}_{qd})^2+\frac98(c^{(1)}_{qu}+c^{(1)}_{qd})^2+\frac{9}{8}(c^{(1)}_{qu}-c^{(1)}_{qd})^2 \, ,
\eea
being the coefficients $c_i$  defined according to \eq{dim6}.
This extends the results of \cite{Chivukula:1996yr}.
It is important to remark that  in \eq{bsmjj} we have included  terms  of 
order $c_i^2/\Lambda^4$;  for $c_i\gg 1$
these terms are enhanced by an extra $c_i$ factor 
with respect to the interference terms (of order $c_i/\Lambda^2$), 
compensating for the  $ \hat s/\Lambda^2$ suppression factor.
Therefore they should  be considered in the calculations.  
Contributions from operators of dimension eight or larger are always   smaller.
For example, 
dimension-eight operators contributing to dijets  involve at most
four-fermions  and extra derivatives, e.g. $\partial^\nu\partial_\nu\bar \psi\gamma^\mu \psi
\bar \psi\gamma_\mu \psi$,
and therefore  their coefficients in front  are not parametrically  larger than those
of dimension-six four-quark operators.
They are then always suppressed by an extra $\sim \hat s/\Lambda^2$.

\begin{figure}[top]
\centering
\hspace{-0.5cm}
\includegraphics[scale=.8]{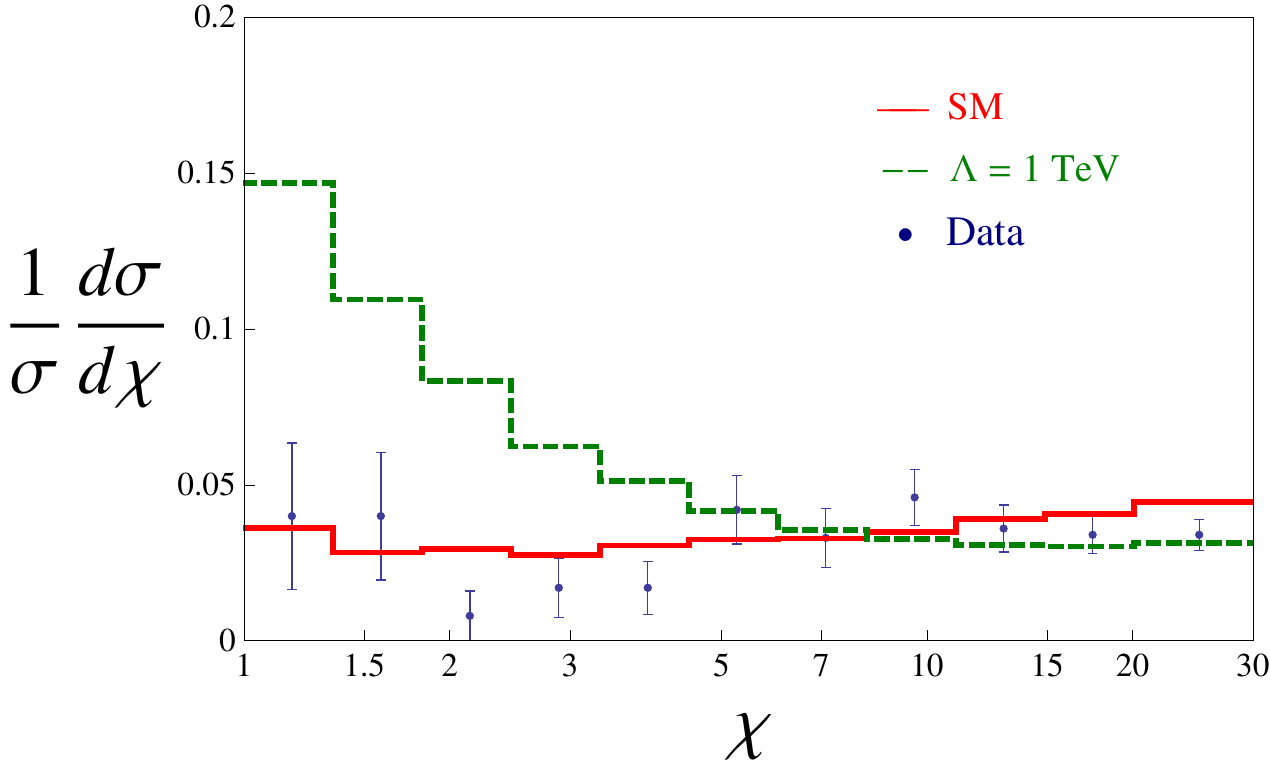} 
\caption{\footnotesize 
Dijet differential cross section
as a function of $\chi$ for $m_{jj}>2$ TeV at the LHC with $\sqrt{s}=7$ TeV.
The QCD contribution is shown in solid red line,
 while the green dashed line includes the contribution from the operator 
 ${\cal O}^{(8)}_{qq}$ with $c^{(8)}_{qq}=-0.5$ and
 $\Lambda=1$ TeV.}
\label{chi_distribution}
\end{figure}

As compared to the SM contribution \eq{smjj}, the  BSM contribution \eq{bsmjj} 
is enhanced at  large $\hat s$  and large CM scattering  angle $\theta^*$, 
or equivalently,   for  large (negative) $\hat{t}=-\hat{s}(1-\cos\theta^*)/2$.
It is convenient to define   the angular variable 
$\chi=(1+|\cos\theta^*|)/(1-|\cos\theta^*|)=-(1+\hat s/\hat t) \in [1,+\infty)$
that can also be written as $\chi=e^{|y_1-y_2|}$
where $y_{1,2}$ are the rapidity of the two jets.
The QCD contribution to the differential cross section
$d\sigma(pp\rightarrow jj)/d\chi$ is almost flat in $\chi$,
 while that of  BSM  grows for small values of $\chi$, as can be appreciated in \fig{chi_distribution}.

\subsection{The   $F_\chi$ parameter}
\label{fchival}

To put bounds on  BSM four-quark operators,
we will follow the method used by the ATLAS collaboration \cite{Collaboration:2010eza,Aad:2011aj}. 
This is based on the variable
$F_\chi$  defined as the 
quotient of events  with $1 < \chi < \chi_c \equiv 3.32$, the central region in the detector, over those with $1<\chi < \chi_{max} \equiv 30$:
\beq
F_\chi^{m_{jj}^{cut}} = \frac{N (\chi < \chi_c, m_{jj} > m_{jj}^{cut})}{N (\chi < \chi_{max},  m_{jj} > m_{jj}^{cut})}\, ,
\label{fchii}
\eeq
where $m_{jj}^{cut}$ is the cut over the invariant mass of the two-jet pair.
Many systematic effects cancel in this ratio, 
providing an optimal test of QCD and a sensitive probe of hard new physics. 
It is also useful to write the analytic expression for this observable.
Defining the integrated differential cross section over the angular region from 1 to $\chi_0$ as
\beq
\sigma^{m_{jj}^{cut}}_{\chi_0} \equiv \int^{\chi_0}_1d\chi \left. \frac{d\sigma}{d\chi} \right|^{m_{jj}^{cut}} _{pp\rightarrow jj}\, ,
\eeq
we have
\beq
F_\chi^{m_{jj}^{cut}} =\frac{\sigma^{m_{jj}^{cut}}_{\chi_c}}{ \sigma^{m_{jj}^{cut}}_{\chi_{max}}} \simeq \frac{(\sigma^{m_{jj}^{cut}}_{\chi_c})_{SM} + (\sigma^{m_{jj}^{cut}}_{\chi_c})_{BSM}}{(\sigma^{m_{jj}^{cut}}_{\chi_{max}})_{SM}}\, ,
\label{fchiap}
\eeq
where we have split the contribution of the SM  from that of the BSM, 
and considered  that the SM contribution, being almost flat, dominates in the denominator. By making this approximation the deviation from the exact value of $F_\chi$ is of order 10\%.
Using \eqs{smjj} and (\ref{bsmjj}),  and   performing the integration over $\chi$, 
we obtain the result
\begin{align}
F^{m_{jj}^{cut}}_\chi&\simeq (F_{\chi}^{m_{jj}^{cut}})_{SM}-\frac{1}{\Lambda^2}\vec{A}\cdot\vec{\cal P}+
\frac{1}{\Lambda^4}\vec{B}\cdot\vec{\cal Q}\, , 
\label{fchi}
\end{align}
where
\begin{align}
\nonumber
\vec{A}&=(A_1^{u},A_2^{u},A_1^{d},A_2^{d},A_3,A_4)\, ,\\
\vec{B}&=(B_1^{u},B_2^{u},B_1^{d},B_2^{d},B_3,B_4)\, ,
\end{align}
and
\begin{eqnarray}
\nonumber
\vec{\cal P}& \simeq &(
0.36
{ P_{uu}},
0.12
{ P_{uu}},
0.36
{ P_{dd}},
0.12
{ P_{dd}},
0.17
{ P_{ud}},
0.074
{ P_{ud}})\, \mathrm{ TeV^2}\, , \\
\nonumber
\vec{\cal Q}& \simeq &(0.013
{ Q_{uu}},
0.0069
{ Q_{uu}},
0.013
{ Q_{dd}},
0.0069
{ Q_{dd}},
0.0024
{ Q_{ud}},
0.00097
{ Q_{ud}})\, \mathrm{ TeV^4}\, ,
\label{vecPQ}
\end{eqnarray}
where $(F_{\chi}^{m_{jj}^{cut}})_{SM}$ is the SM value of $F_\chi^{m_{jj}^{cut}}$ and the coefficients $P_{q_i q_j}$
and  $Q_{q_i q_j}$ encode the integration over the parton distribution functions (PDF):
\begin{eqnarray}
{P_{q_i q_j}}&=&\frac{1}{(\sigma^{m_{jj}^{cut}}_{\chi_{max}})_{SM}} 
\int d\tau \! \int dx \, f_{q_i}(x) f_{q_j}(\tau/x) \frac{\alpha_s(\tau s)}{x} + (i \leftrightarrow j \ \mathrm{for} \ i \neq j)
\, ,\\
{ Q_{q_i q_j}}&=&\frac{1}{(\sigma^{m_{jj}^{cut}}_{\chi_{max}})_{SM}} 
\int d\tau \! \int dx \, f_{q_i}(x) f_{q_j}(\tau/x) \frac{\tau s}{x}
+ (i \leftrightarrow j \ \mathrm{for} \ i \neq j)
\, ,
\end{eqnarray}
where $\hat{s} = \tau s$ is the center of mass energy of the partons $q_i q_j$ that initiate the collision, with $\sqrt{s}=7$ TeV the center of mass energy of the colliding protons.
To calculate these coefficients we use
MadGraph/MadEvent 4.4.57 \cite{Alwall:2007st} and implement the cuts taken in \cite{Aad:2011aj}.
For this analysis we use the CTEQ6L1 PDF set and fix both the renormalization and factorization scales to $m_{jj}^{cut} = 2 \TeV$. 
We obtain:
\begin{eqnarray}
&& {P_{uu}}\simeq 0.23 \, \mathrm{ TeV^2}\ , \ \ {P_{dd}}\simeq 0.038 \, \mathrm{ TeV^2}\ ,  \ \ {P_{ud}} \simeq 0.28 \, \mathrm{ TeV^2}\, ,\nonumber \\ 
&& {Q_{uu}}\simeq 23 \, \mathrm{ TeV^4}\ ,\ \ {Q_{dd}}\simeq 3.8 \, \mathrm{ TeV^4}\  ,\ \ {Q_{ud}}\simeq 19 \, \mathrm{ TeV^4}\, , 
\label{pandq}
\end{eqnarray}
and $(F_{\chi}^{\, 2\TeV})_{SM}\simeq 0.067$, $(\sigma^{\, 2\TeV}_{\chi_{max}})_{SM} \simeq 0.016 \TeV^{-2}$.
We have checked the consistency of these results by numerically integrating over the MSTW2008 PDFs \cite{Martin:2009iq} using our analytical formulae for the cross sections \eqs{smjj} and (\ref{bsmjj})
and implemeting the cuts in \cite{Aad:2011aj}, which translate into the integration limits $x \in [\sqrt{\tau} e^{-y_B^{cut}},1], \tau \in [(m_{jj}^{cut})^2/s,e^{-2y_B^{cut}}]$ and $x \in [\tau,1], \tau \in [e^{-2y_B^{cut}},1]$, where $y_B^{cut} = 1.1$ is the cut on the rapidity boost of the partonic center of mass, $|y_B| = {1\over 2} |y_1 + y_2| \leqslant y_B^{cut}$.
The variation of renormalization and factorization scales (by twice and half) introduces a theoretical uncertainty of the order of $10-15\%$. 
We have not computed the errors arising from   the PDFs, and have not taken into account hadronization or showering effects since it is reasonable to neglect them for high dijet invariant masses \cite{Perelstein:2010hh}.

\section{Bounds}
\label{bounds}

ATLAS has reported  angular distributions of  dijets for several $m_{jj}$ \cite{Aad:2011aj}. 
We are interested in those with the  largest invariant masses  that correspond to 
$m_{jj}>2 \TeV$ and are given in \fig{chi_distribution}. Using this data and 
\eq{fchii}, we obtain $F_\chi^{\, 2\TeV}=0.053\pm 0.015$,
and therefore the 95$\%$ CL bound
\beq
0.023<F_\chi^{\, 2\TeV}<0.083\, .
\label{exp}
\eeq
Using this value and the prediction \eq{fchi}  
one can obtain  bounds on the  scales  suppressing the operators \eq{4u}.
We instead derive the bounds using the coefficients of \eq{pandq} but without making any approximation on the denominator of \eq{fchiap}.
The results are shown in Table~\ref{opbounds}. 
\begin{table}[top]
	\begin{center}
\begin{tabular}{ccc}
\hline
Operator & $\Lambda_- (\TeV)$ & $\Lambda_+ (\TeV)$  \\
\vspace{-0.5cm} \\
\hline \\
\vspace{-1cm} \\
$ \Op_{uu}^{(1)}$ & $ 3.2 $ & $ 2.1 $  \\
$ \Op_{dd}^{(1)}$ & $ 1.8 $ & $ 1.5 $  \\
$ \Op_{ud}^{(1)}$ & $ 1.5 $ & $ 1.5 $ \\
$ \Op_{ud}^{(8)}$ & $ 1.3 $ & $ 0.8 $  \\
$ \Op_{qq}^{(1)}$ & $ 3.5 $ & $ 2.4 $ \\
$ \Op_{qq}^{(8)}$ & $ 2.5 $ & $ 1.3 $  \\
$ \Op_{qu}^{(1)}$ & $ 1.7 $ & $ 1.7 $ \\
$ \Op_{qu}^{(8)}$ & $ 1.4 $ & $ 1.0 $ \\
$ \Op_{qd}^{(1)}$ & $ 1.3 $ & $ 1.3 $  \\
$ \Op_{qd}^{(8)}$ & $ 1.0 $ & $ 0.8 $  \\
\hline
\end{tabular} \\
	\caption{Bounds at 95\% CL on the  scale suppressing the four-quark interactions. 
	We  denote by $\Lambda_{\pm}$ the bound on this scale   obtained when taking the coefficient
	in front of the operator $c_i=\pm 1$, and considering  the effects of the operators one by one.} 
	\label{opbounds}
	\end{center}
\end{table} 
Few comments are in order.  
These bounds are obtained by taking the coefficient in front of the corresponding operator to be $c_i= \pm 1$.
For other values we must rescale the bound by a factor $\sqrt{|c_i|}$.
Since  we are working in the 
 approximation  in which  the energy of the physical process  is assumed 
 to be smaller  than  the masses  of the BSM  states $\Lambda$,  
 we must require $\Lambda> m_{jj}^{cut}$.
This implies that our bounds can only   strictly  be applied if
 $|c_i|> (2\ \TeV/\Lambda_\pm)^2$.
We recall that large values  of $c_i$ are in principle possible since 
$c_i\lesssim 16\pi^2$.
Also, we notice that for $c_i<0$
the interference between the BSM contribution and the  QCD   contribution is constructive
(with the exception of  ${\cal O}_{ud,qu,qd}^{(1)}$ where the interference is null),
and as a consequence the bound is stronger than  for a positive $c_i$.

These bounds are subject to a set of theoretical errors. 
The uncertainty in the parameters \eq{pandq} estimated by changing the factorization and renormalization scales results in a $\sim 5 \%$ uncertainty in the bounds.
Also the  NLO QCD correction to $(F_{\chi}^{\, 2 \TeV})_{BSM}/(F_{\chi}^{\, 2 \TeV})_{SM}$ has shown to be as large as   $\sim 30 \%$ \cite{Gao:2011ha}, what amounts to a $\sim 10 \%$ uncertainty in the bounds on $\Lambda$. Finally, it has been recently shown in \cite{arXiv:1201.3926} that electroweak corrections reduce the SM prediction of $F_\chi$ by a $\sim 2 \%$ for large invariant masses $m_{jj} \sim 2 \TeV$.
We therefore expect that our calculations for the bounds on $\Lambda$
 can be trusted within a $\sim10\%$ margin of error.

\subsection{Bounds on composite quarks}

As we mentioned in section~\ref{sec1},
previous experiments   have not  been able to probe the 
compositeness of quarks
beyond the TeV scale.
Data from dijets at the LHC  can however improve 
this situation and put  stronger  constraints on their compositeness scale.

We will focus  on models in which quarks arise as composite states of 
a strong sector whose global symmetry is
${\cal G}\equiv SU(3)_c\otimes SU(2)_L\otimes U(1)_Y\otimes G_F$,
where $G_F$ is given in \eq{gf}.
In these theories we expect to have massive vector resonances associated to the current operators of  $\cal G$, and then  transforming  in the adjoint representation
of ${\cal G}$.
This is in fact the case of the  five-dimensional analogs  based on the AdS/CFT correspondence \cite{Agashe:2004rs}.
Following  Ref.~\cite{Giudice:2007fh}, we will 
assume that all the vector resonances  have equal  masses and couplings,  $m_\rho$ and   $g_\rho$ respectively.
Let us first  consider  the case in which only   the right-handed up-type quarks $u_R$ are  composite states, with 
charges under the global group $\cal G$  equal  to $\bf (3,1,2/3,1,1,2)$.
In this type of models, as we said before, the Higgs could also be composite  without affecting our conclusions.
Now, integrating out the heavy vector resonances at tree level (an approximation valid  in the large-$N$ limit or, equivalently, $g_\rho \ll 4 \pi$), we find that the  four-quark operators of \eq{4u} are induced with coefficients given in Table~\ref{coefs},
where we have fixed $\Lambda= m_\rho$.  
Constraints
from  dijets give  $f\equiv m_\rho/g_\rho\gtrsim 2$ TeV.
For $g_\rho\gg 1$,  we see that this bound is stronger than  that coming from
 the $S$-parameter   that requires  $f\gtrsim  4\pi v/g_\rho$  in theories of composite Higgs \cite{Giudice:2007fh}.

\begin{table}[top]
	\begin{center}
\begin{tabular}{ccccccccccc}
\hline
Composite & $c_{uu}^{(1)}/g_\rho^2$ & $c_{dd}^{(1)}/g_\rho^2$ & $c_{ud}^{(1)}/g_\rho^2$ & $c_{ud}^{(8)}/g_\rho^2$ & $c_{qq}^{(1)}/g_\rho^2$ & $c_{qq}^{(8)}/g_\rho^2$ & $c_{qu}^{(1)}/g_\rho^2$ & $c_{qu}^{(8)}/g_\rho^2$ & $c_{qd}^{(1)}/g_\rho^2$ & $c_{qd}^{(8)}/g_\rho^2$ \\
\vspace{-0.5cm} \\
\hline \\
\vspace{-1cm} \\
$u_R$ & $ -37/72 $ & $ 0 $ & $ 0 $ & $ 0 $ & $ 0 $ & $ 0 $ & $ 0 $ & $ 0 $ & $ 0 $ & $ 0 $  \\
$d_R$ & $ 0 $ & $ -7/18 $ & $ 0 $ & $ 0 $ & $ 0 $ & $ 0 $ & $ 0 $ & $ 0 $ & $ 0 $ & $ 0 $ \\
$u_R, d_R$ & $ -37/72 $ & $ -7/18 $ & $ 2/9 $ & $ -1 $ & $ 0 $ & $ 0 $ & $ 0 $ & $ 0 $ & $ 0 $ & $ 0 $  \\
$q_L$ & $ 0 $ & $ 0 $ & $ 0 $ & $ 0 $ & $ -5/36 $ & $ -1 $ & $ 0 $ & $ 0 $ & $ 0 $ & $ 0 $\\
$q_L, u_R$ & $ -37/72 $ & $ 0 $ & $ 0 $ & $ 0 $ & $ -5/36 $ & $ -1 $ & $ -1/9 $ & $ -1 $ & $ 0 $ & $ 0 $\\
$q_L, d_R$ & $ 0 $ & $ -7/18 $ & $ 0 $ & $ 0 $ & $ -5/36 $ & $ -1 $ & $ 0 $ & $ 0 $ & $ 1/18 $ & $ -1 $\\
$q_L, u_R, d_R$ & $ -37/72 $ & $ -7/18 $ & $ 2/9 $ & $ -1 $ & $ -5/36 $ & $ -1 $ & $ -1/9 $ & $ -1 $ & $ 1/18 $ & $ -1 $ \\
\hline
\end{tabular} \\
	\caption{Coefficients of the operators of \eq{4u}   induced from  integrating out  heavy vector resonances for different composite quark scenarios. We have taken $\Lambda= m_\rho$.}
	\label{coefs}
	\end{center}
\end{table}

 \begin{table}[top]
	\begin{center}
\begin{tabular}{cc}
\hline
Composite States & $f\, (\TeV)$ \\
\vspace{-0.5cm} \\
\hline \\
\vspace{-1cm} \\
$d_R$ & $ 1.1 $  \\
$u_R$ & $ 2.3 $  \\
$u_R, d_R$ & $ 2.6 $  \\
$q_L$ & $ 2.7 $\\
$q_L, d_R$ & $ 2.9 $\\
$q_L, u_R$ & $ 3.5 $\\
$q_L, u_R, d_R$ & $ 3.8 $  \\
\hline
\end{tabular} \\
	\caption{ 95\% CL bounds on the scale $f=m_\rho/g_\rho$ for different composite quark scenarios.}
	\label{scebounds}
	\end{center}
\end{table}

 Similarly, we can assume a scenario 
 where only the right-handed down-type quarks $d_R$ are composite with quantum
 numbers under $\cal G$ equal to  $\bf (3,1,-1/3,1,3,1)$.
 Again
the coefficients of the  four-quark operators  induced are given in Table~\ref{coefs}.
We obtain the bound $f\gtrsim 1$ TeV.
In the case of  both $u_R$ and $d_R$ composite the bound goes up to
$f\gtrsim 2.5$ TeV.

For composite left-handed quarks $q_L$   with $\cal G$-charges $\bf (3,2,1/6,3,1,1)$, the bound is $f\gtrsim 3$ TeV.
Bounds on other composite quark scenarios are  given  in Table~\ref{scebounds}.

For weakly-coupled  resonances ($g_\rho\lesssim 1$)  with masses  close to $m_{jj}^{cut}$  stronger bounds  can be obtained from dijet resonance searches at the LHC  \cite{Redi:2011zi}. 
This is just a consequence of the resonant enhancement of the cross section for a narrow region of invariant masses, where the resonances sit. This feature however is lost when the resonances are too broad.

\subsection{Bounds on heavy gauge bosons}
\label{kkglu}

\begin{figure}[t!]
\centering
\hspace{-0.5cm}
\includegraphics[scale=.9]{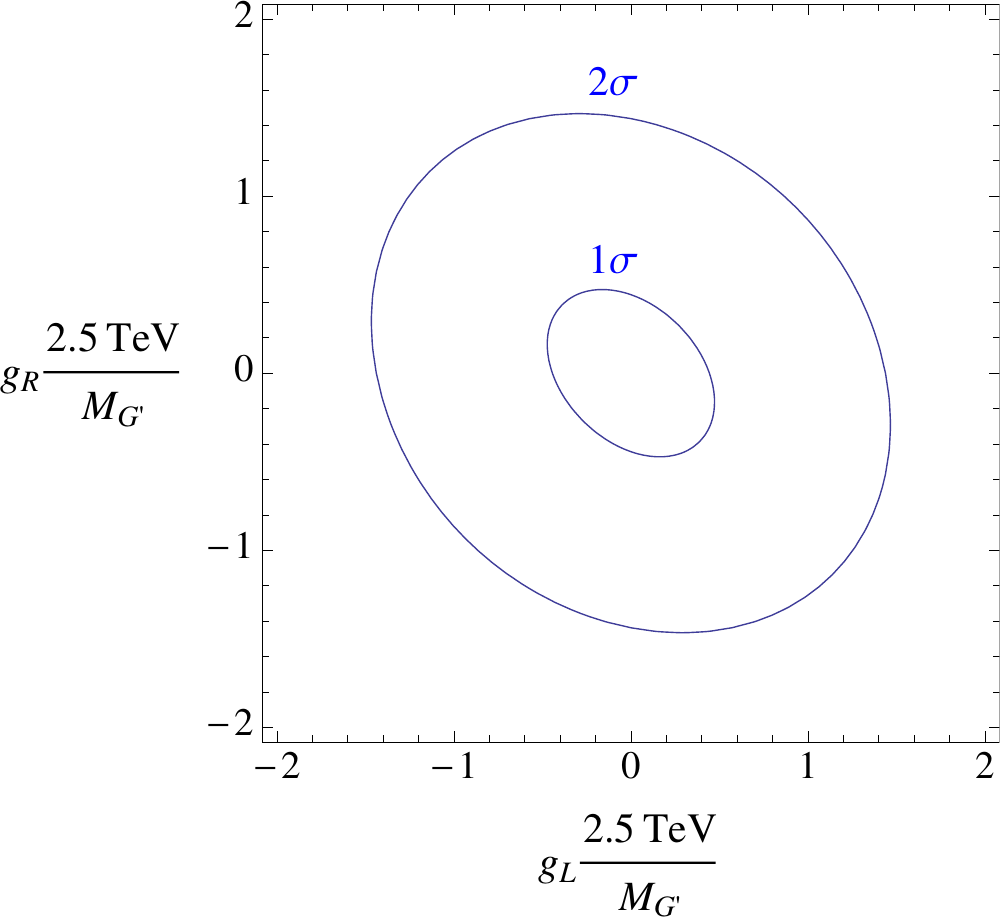} 
\caption{\footnotesize 
Excluded  region in the $g_L-g_R$ plane by the $m_{jj}>2$ TeV dijet  analysis.}
\label{KKgluon}
\end{figure}

Heavy gauge  bosons at the TeV-scale  coupled to first family quarks  generate four-quark
operators that can be constrained by the dijet LHC data. 
Here we provide some examples.
For a gauge boson $Z'$ gauging  baryon number or
  hypercharge  we obtain respectively
\beq
\frac{M_{Z'_B}}{g_B}\gtrsim 1.2 \TeV\ , \ \ \ \ 
\frac{M_{Z'_Y}}{g_Y}\gtrsim 1.6 \TeV\, ,
\label{z'}
\eeq
while for the gauge bosons $W'$  of  a $SU(2)_R$ symmetry, where   $q_{R}=(u_{R},d_{R})$ is assumed to transform as a 
doublet, we get
\beq
\frac{M_{W'}}{g_R}\gtrsim 1.6 \TeV\, .
\label{w'}
\eeq
As mentioned before, the fact that we work within an effective theory  \eq{dim6}, 
implies  that  our bounds  only  apply   to  resonances with masses above    $m_{jj}^{cut}=2$ TeV.
Gluonic resonances $G'^A_\mu$ coupled to first family quarks as 
$
{\cal L}_{int}=G'^A_\mu\left[g_L \bar q_L T^A \gamma^\mu q_L +g_R \bar q_R T^A \gamma^\mu q_R \right]
$,
with  $T_A = \lambda_A/2$ where $\lambda_A$ are the Gell-Mann matrices,
can also be constrained.
This kind of resonances have been recently advocated (see for example Ref.~\cite{arXiv:1107.1473})
to accommodate the discrepancy in the top forward-backward asymmetry measured at Tevatron.
In \fig{KKgluon} we show the excluded region of the parameter space. It can be seen that, for a resonance of mass $M_{G'} = 2.5$ TeV, the allowed range for the couplings is
 $-1.5\lesssim g_{L,R}\lesssim 1.5$  at 95\% CL.
Similar   bounds   have been also obtained in  
 Ref.~\cite{Haisch:2011up}.
 
\subsection{Bounds on oblique parameters $Y$, $W$ and $Z$}
\label{oblique}

\begin{figure}[t!]
\centering
\hspace{-0.5cm}
\includegraphics[scale=.9]{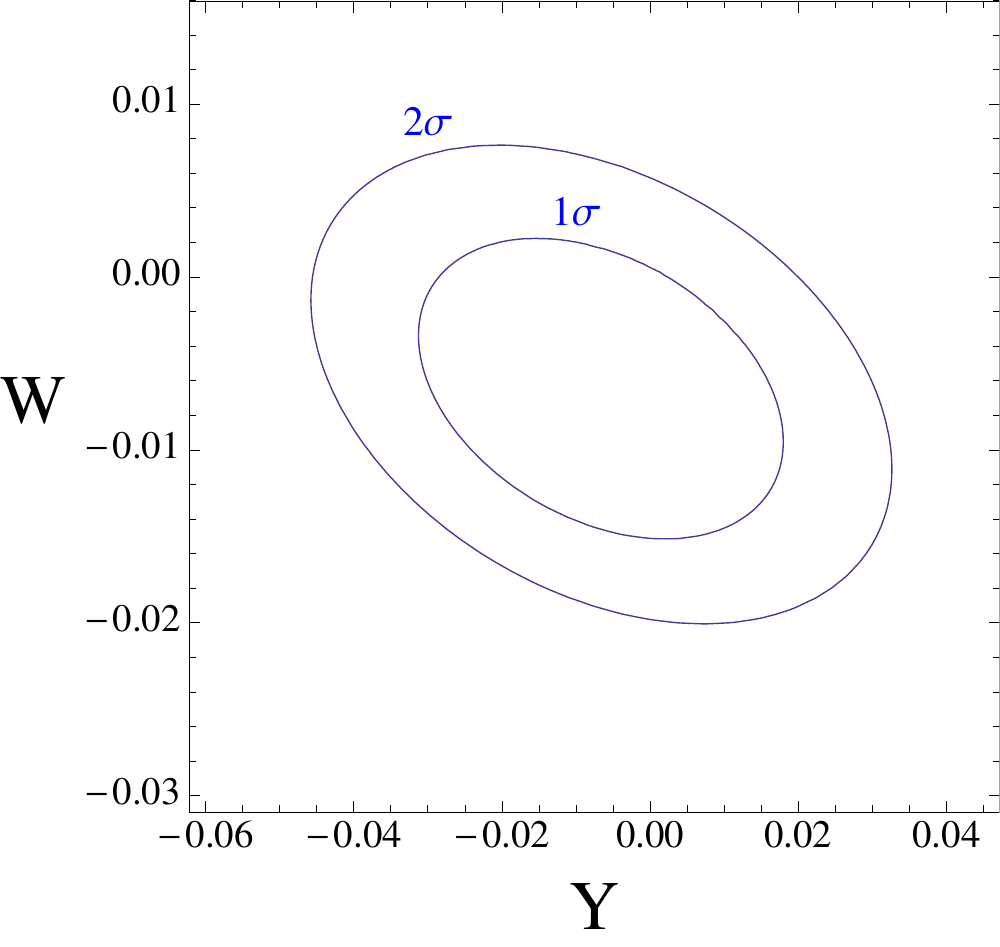} 
\caption{\footnotesize 
Excluded  region in the $W$-$Y$ plane by the $m_{jj}>2$ TeV ATLAS dijet analysis.}
\label{WY}
\end{figure}

The electroweak precision parameters $Y$, $W$ and $Z$ \cite{Barbieri:2004qk}
can be regarded as a measure of the compositeness of the transversal components of the $SU(2)_L$, $U(1)_Y$, and $SU(3)_c$ gauge bosons respectively.
They manifest themselves as deviations of the self-energies of such vector bosons, 
and can be parametrized by the following higher dimensional operators:
\beq
\frac{-Y}{4 m_W^2} (\partial_\rho B_{\mu\nu})^2, \quad \frac{-W}{4 m_W^2} (D_\rho W_{\mu\nu}^I)^2, \quad \frac{-Z}{4 m_W^2} (D_\rho G_{\mu\nu}^{A})^2\, .
\eeq
At large momenta as compared to the masses of the gauge bosons, these operators induce effective four-fermion operators, equivalent to those arising from integrating out a very heavy copy of the corresponding gauge boson. 
Therefore our dijet analysis can be conveniently used to put bounds on these parameters.
We show in \fig{WY} our results in the $W$-$Y$ plane. Although  bounds from LEP 
\cite{Barbieri:2004qk} are still stronger,  this analysis shows that LHC will be competitive 
when  running at a higher energy.
Regarding the $Z$-parameter our analysis gives the strongest bound up to date:
\beq
-3 \times 10^{-3} \lesssim Z \lesssim 6 \times 10^{-4}.
\label{Zparameter}
\eeq

\subsection{Bounds on   new  interactions for the  $A_{FB}$ of the top}
\label{Afb}

The recent discrepancy between the measured $A_{FB}$ of the top and its SM prediction 
\cite{Aaltonen:2011kc, Abazov:2011rq}
has boosted  the search for BSM  that could explain it.
Dijet  angular distributions can be  useful  to constrain these models.
As an example, we consider the proposal of  
Refs.~\cite{Degrande:2010kt,arXiv:1103.2297} where the measured value of the top asymmetry was  explained by the following new interaction:
\bea
{\cal L}_{eff}= \frac{c^{(8)}_A}{\Lambda^2}\Op^{(8)}_{A}=\frac{c^{(8)}_A}{\Lambda^2}(\bar{u}\,T^A\gamma^\mu\gamma^5u)(\bar{t}\,T^A\gamma_\mu\gamma^5t)\, .
\eea
In terms of chirality eigenstates the operator $\Op^{(8)}_{A}$ reads
\begin{align}
\nonumber
\Op^{(8)}_{A}=&(\bar{u}_R\gamma^\mu T^A u_R)(\bar{t}_R\gamma_\mu T^A t_R)-(\bar{u}_L\gamma^\mu T^A u_L)(\bar{t}_R\gamma_\mu T^A t_R)\\
&-(\bar{u}_R\gamma^\mu T^A u_R)(\bar{t}_L\gamma_\mu T^A t_L)+(\bar{u}_L\gamma^\mu T^A u_L)(\bar{t}_L\gamma_\mu T^A t_L)\, .
\label{newope}
\end{align}
If these operators arise from BSM  that are invariant under the SM gauge group and $G_F$  (up to small effects $v^2/\Lambda^2$ and Yukawa couplings),
the presence of $c^{(8)}_A\not=0$ requires, in the  basis  \ref{4fer}, 
\bea
c^{(8)}_{ut} = -c^{(8)}_{qt} = -c^{(8)}_{qu} = c^{(8)}_{qq} = c^{(8)}_A\, .
\eea
In other words,  the flavor symmetry 
requires that if the operator $\Op^{(8)}_{A}$ is generated, also operators involving four 
up-quarks must be present.
Bounds from our dijet analysis (mostly from the bounds on $c^{(8)}_{qu}$  and 
$c^{(8)}_{qq}$) lead then to
\beq
\frac{c^{(8)}_A}{\Lambda^2}\lesssim \frac{0.4 }{\TeV^{2}}\, ,
\label{Afb_flav}
\eeq
 excluding the possibility to fit  the recent  top asymmetry  measurement
 which requires  $c^{(8)}_A/\Lambda^2\sim 2$ TeV$^{-2}$ \cite{arXiv:1103.2297}.

\begin{figure}[t!]
\centering
\hspace{-0.5cm}
\includegraphics[scale=.9]{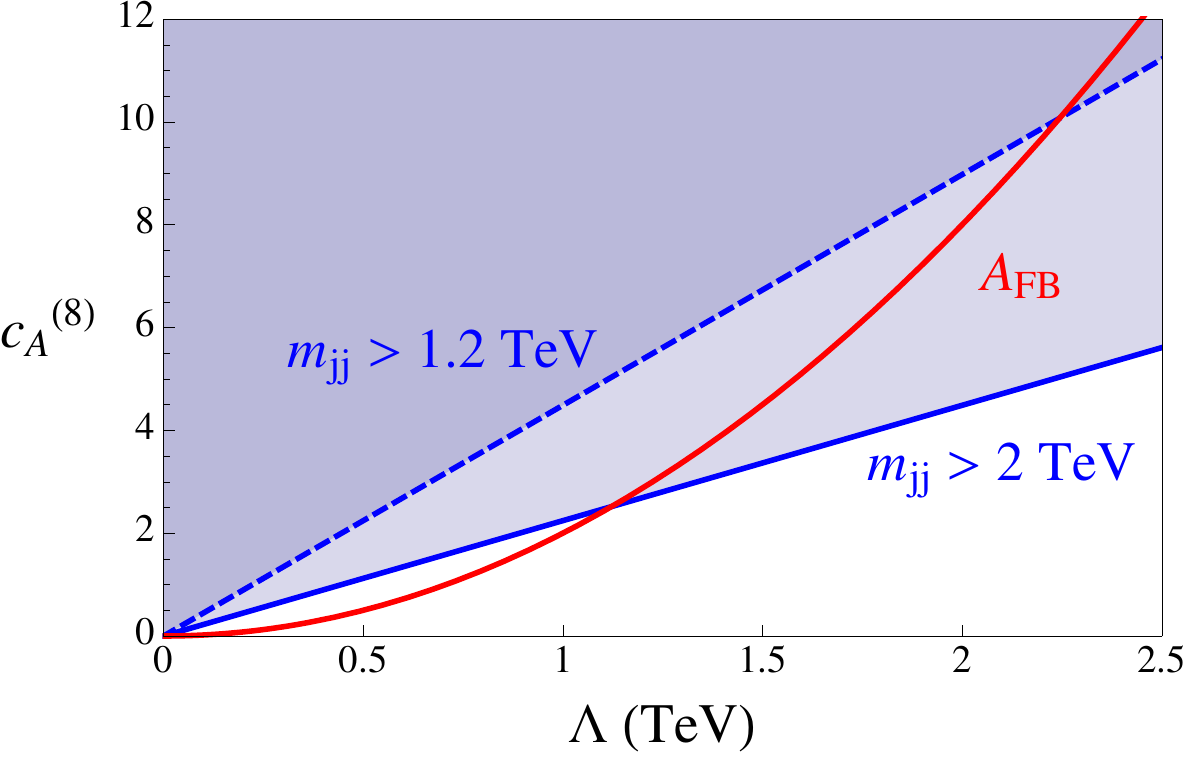} 
\caption{\footnotesize 
The red line shows the value of $c_A^{(8)}$ as a function of $\Lambda$  that fits the $A_{FB}$ of the top \cite{arXiv:1103.2297}.
The shaded regions delimited by the solid and dashed blue lines show the excluded region due to our
dijet  angular distribution analysis  with cuts $m_{jj}^{cut} = 2 \TeV$ and $m_{jj}^{cut} = 1.2 \TeV$ respectively. 
}
\label{AFB}
\end{figure}

If we relax the assumption of flavor invariance of the BSM sector,
an operator involving four up-quarks can still be generated from 
$\Op^{(8)}_A$ at the one-loop level.
The one-loop contribution, involving tops, is divergent  and therefore  sensitive to physics at  the BSM scale $\Lambda$.
We can get an estimate by regulating the divergence  with a hard cut-off taken to be $\Lambda$.
We obtain 
\bea
c^{(8)}_{qq} = \frac13 c^{(8)}_{uu} = -2c^{(8)}_{qu} \simeq  -\frac{(c^{(8)}_A)^2}{4\pi^2}\, .
\label{gene}
\eea
In Fig.~\ref{AFB} we show the region of the parameter space that fits the $A_{FB}$ of the top  with the  region excluded by dijets using \eq{gene}. 
One can see that   dijets with $m_{jj}>2$ TeV  exclude a  large region of the 
parameter space, although, as we mentioned before,    
 these results cannot be strictly  applied  if $\Lambda<m_{jj}$.  For this reason we also show the exclusion region arising from dijets
with  smaller invariant masses, $m_{jj}>1.2$ TeV.

\section{Conclusions}

We have shown that the present dijet LHC data is  already testing all the quark sector of the SM
 at almost  the same level  of accuracy as  leptons were tested at LEP after a decade  of  data collection. 
This is  due to the  very high-energy  of the dijet scattering at the LHC
that enhances  the effects from four-quark interactions.
We have used the $F_\chi$ parameter, 
defined in \eq{fchii},
 to put bounds  
on all possible four-quark operators.
Our results,   presented in  Table~\ref{opbounds}, show
 bounds  on  the scales suppressing these operators 
 $\Lambda/\sqrt{|c_i|}$   ranging between  $1-3$ TeV.

Among the most interesting BSM scenarios to be tested by  dijet  angular distributions
are  theories in which  strong dynamics are postulated to solve the hierarchy problem.
These theories  demand  a  composite scale   $\Lambda$ around the TeV-scale.
We have seen  that if  the SM quarks are  composite states arising
from a new  strong sector,        $\Lambda\gtrsim 50\ \TeV\times (g_\rho/4\pi)$. 
Other possibilities are also significantly constrained,  as can be seen from  Table~\ref{scebounds}.
We also derive the best bounds on  the  $Z$-parameter, \eq{Zparameter},  that  measures  the 
degree of compositeness of  the gluons.

We also show that extra gauge bosons with sizable couplings to quarks are  constrained to lie above the TeV scale,  limiting  then their possible contribution to  the  $A_{FB}$ of the top.

Finally, we would like to stress that these results are based on the 2010 LHC data corresponding to 36 pb$^{-1}$ of integrated luminosity \cite{Aad:2011aj}. It is expected that the 2011 LHC data set, containing more luminosity,  will significantly improve all the bounds derived throughout this analysis. 


\vspace{1cm}

\textbf{Note Added:}  The 2011 data set for dijet events at CMS has been recently reported in Ref.~\cite{Chatrchyan:2012bf}, corresponding to a luminosity of $2.2$ fb$^{-1}$ and a cut in the dijet invariant mass of $m_{jj}>3$ TeV. The analysis made throughout this article can be applied in this case; by doing so we obtain more stringent bounds in all our results. 

In this new analysis we define the observable $F_\chi$ as the 
quotient of events  with $1 < \chi < \chi_c \equiv 3$, the first bin in the experimental analysis, over those with $1<\chi < \chi_{max} \equiv 16$:
\beq
F_\chi^{m_{jj}^{cut}} = \frac{N (\chi < \chi_c, m_{jj} > m_{jj}^{cut})}{N (\chi < \chi_{max},  m_{jj} > m_{jj}^{cut})}\, .
\label{fchi}
\eeq
The experimental value for this observable is $F_\chi^{(m_{jj}>3\text{ TeV})} \simeq 0.09$ with a 2$\sigma$ interval,
\bea
0.003 \lesssim F_\chi^{(m_{jj}>3\text{ TeV})} \lesssim 0.15\,\,\,\,\,\,\, \text{at 95\% C.L. },
\eea
while the SM prediction is $F_\chi^{(m_{jj}>3\text{ TeV})} \simeq 0.12$. The new definition of $F_\chi$ toghether with the new cut in the dijet invariant mass will modify the numerical results in Sec. \ref{fchival} in the following way: first of all the values in \eq{vecPQ} and \eq{pandq} have to be replaced by the following ones,
\begin{eqnarray}
\nonumber
\vec{\cal P}& \simeq & \frac{1}{(\sigma^{\, 3\TeV}_{\chi_{max}})_{SM}}(
0.33
{ P_{uu}},
0.10
{ P_{uu}},
0.33
{ P_{dd}},
0.10
{ P_{dd}},
0.15
{ P_{ud}},
0.064
{ P_{ud}})\, \mathrm{ TeV^2}\, , \\
\nonumber
\vec{\cal Q}& \simeq & \frac{1}{(\sigma^{\, 3\TeV}_{\chi_{max}})_{SM}}(0.012
{ Q_{uu}},
0.0064
{ Q_{uu}},
0.012
{ Q_{dd}},
0.0064
{ Q_{dd}},
0.0022
{ Q_{ud}},
0.00087
{ Q_{ud}})\, \mathrm{ TeV^4}\, ,
\end{eqnarray}
and
\begin{eqnarray}
&& {P_{uu}}\simeq 0.013 \, \ , \ \ {P_{dd}}\simeq 0.0019 \, \ ,  \ \ {P_{ud}} \simeq 0.015 \, \, ,\nonumber \\ 
&& {Q_{uu}}\simeq 2.8 \, \mathrm{ TeV^2}\ ,\ \ {Q_{dd}}\simeq 0.37 \, \mathrm{ TeV^2}\  ,\ \ {Q_{ud}}\simeq 2.5 \, \mathrm{ TeV^2}\, , 
\end{eqnarray}
where $(\sigma^{\, 3\TeV}_{\chi_{max}})_{SM} \simeq 0.0131 \TeV^{-2}$.

From these values we can obtain the results corresponding to the new data set. All the bounds derived in Sec.~\ref{bounds} have to be replaced by the ones given in this added note. Concerning the 4-quark operators the results are shown in Table \ref{opboundsup}, while for the composite-quark scenarios the updated bounds are given in Table \ref{sceboundsup}.

\begin{table}[top]
	\begin{center}
\begin{tabular}{cccc}
\hline
Operator & $\Lambda_- /\sqrt{c_i} $ & $\Lambda_+/\sqrt{c_i}$ & $ (\TeV)$  \\
\vspace{-0.5cm} \\
\hline \\
\vspace{-1cm} \\
$ \Op_{uu}^{(1)}$ & $ 4.5 $ & $ 3.0 $  & $ $\\
$ \Op_{dd}^{(1)}$ & $ 2.4 $ & $ 2.0 $  & $ $\\
$ \Op_{ud}^{(1)}$ & $ 2.2 $ & $ 2.2 $ & $ $\\
$ \Op_{ud}^{(8)}$ & $ 1.8 $ & $ 1.3 $  & $ $\\
$ \Op_{qq}^{(1)}$ & $ 5.0 $ & $ 3.5 $ & $ $\\
$ \Op_{qq}^{(8)}$ & $ 3.4 $ & $ 2.0 $  & $ $\\
$ \Op_{qu}^{(1)}$ & $ 2.5 $ & $ 2.5 $ & $ $\\
$ \Op_{qu}^{(8)}$ & $ 1.9 $ & $ 1.5 $ & $ $\\
$ \Op_{qd}^{(1)}$ & $ 1.9 $ & $ 1.9 $  & $ $\\
$ \Op_{qd}^{(8)}$ & $ 1.4 $ & $ 1.2 $  & $ $\\
\hline
\end{tabular} \\
	\caption{Bounds at 95\% CL on the  scale suppressing the four-quark interactions corresponding to the 2011 dijet data set by CMS. 
	We  denote by $\Lambda_{\pm}$ the bound on this scale   obtained when taking the coefficient
	in front of the operator $c_i=\pm 1$, and considering  the effects of the operators one by one.} 
	\label{opboundsup}
	\end{center}
\end{table} 

 \begin{table}[top]
	\begin{center}
\begin{tabular}{cc}
\hline
Composite States & $f\, (\TeV)$ \\
\vspace{-0.5cm} \\
\hline \\
\vspace{-1cm} \\
$d_R$ & $ 1.5 $  \\
$u_R$ & $ 3.2 $  \\
$u_R, d_R$ & $ 3.6 $  \\
$q_L$ & $ 3.8 $\\
$q_L, d_R$ & $ 4.0 $\\
$q_L, u_R$ & $ 4.9 $\\
$q_L, u_R, d_R$ & $ 5.2 $  \\
\hline
\end{tabular} \\
	\caption{ 95\% CL bounds on the scale $f=m_\rho/g_\rho$ for different composite quark scenarios corresponding to the 2011 dijet data set by CMS.}
	\label{sceboundsup}
	\end{center}
\end{table}

\begin{figure}[t!]
\centering
\hspace{-0.5cm}
\includegraphics[scale=.9]{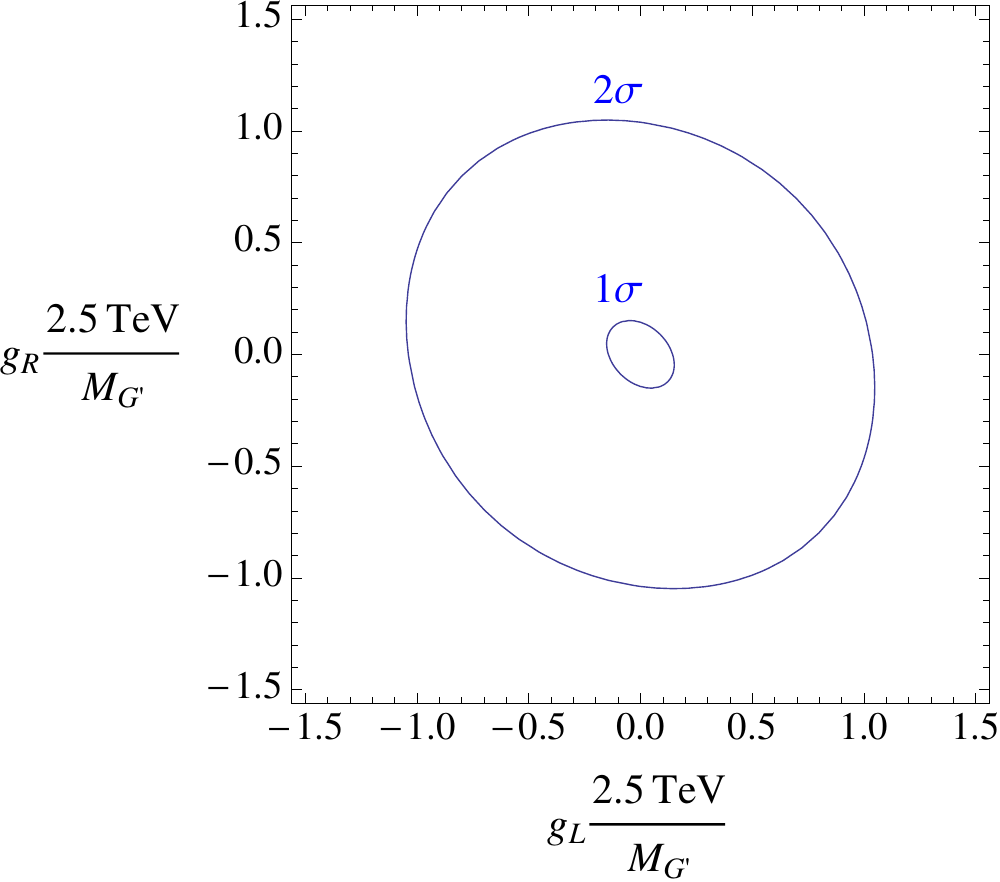} 
\caption{\footnotesize 
Excluded  region in the $g_L-g_R$ plane by the $m_{jj}>3$ TeV CMS dijet  analysis.}
\label{KKgluonup}
\end{figure}

The results given in Sec. \ref{kkglu} have to be replaced by the following ones,
\beq
\frac{M_{W'}}{g_R}\gtrsim 2.3 \TeV\ ,\ \ \ \ \ \ \  
\frac{M_{Z'_B}}{g_B}\gtrsim 1.6 \TeV\ , \ \ \ \ 
\frac{M_{Z'_Y}}{g_Y}\gtrsim 2.3 \TeV\, \ \ \ \  \text{at 95\% C.L.}\, ,
\label{z'}
\eeq
including \fig{KKgluonup} for a gluonic resonance coupled to both LH and RH quarks.

\begin{figure}[t!]
\centering
\hspace{-0.5cm}
\includegraphics[scale=.9]{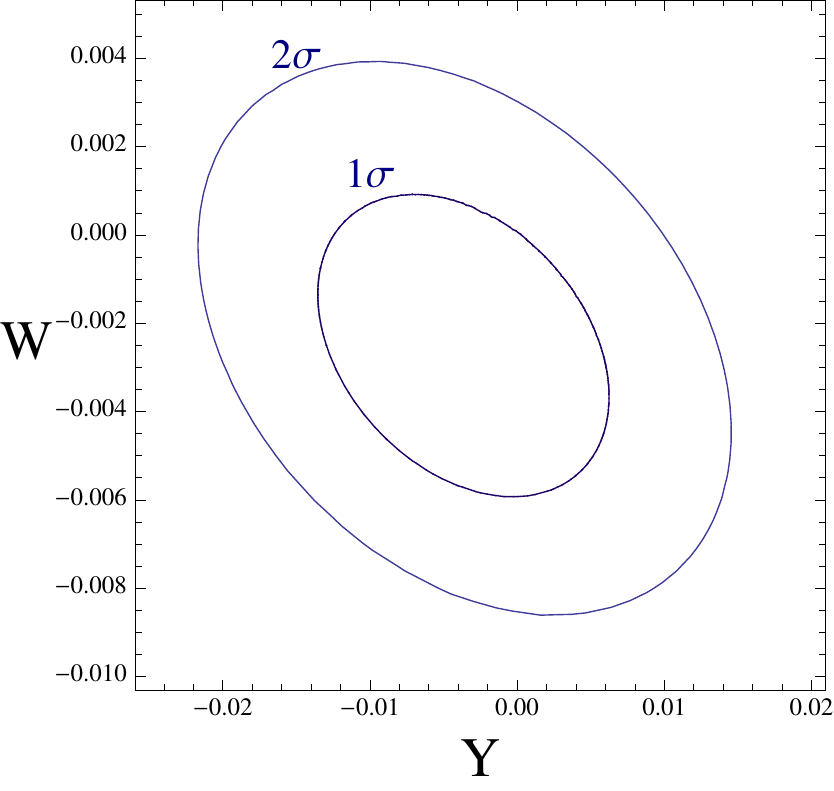} 
\caption{\footnotesize 
Excluded  region in the $W$-$Y$ plane by the $m_{jj}>3$ TeV CMS dijet analysis.}
\label{WYup}
\end{figure}

In the case of Sec.~\ref{oblique} the new bound for the $Z$ parameter is,
\beq
-9 \times 10^{-4} \lesssim Z \lesssim 3 \times 10^{-4}.
\label{Zparameterup}
\eeq 
While the bounds for the $W$ and the $Y$ parameters are shown in \fig{WYup}.

Finally, concerning Sec.~\ref{Afb} and the flavor invariant case we have to replace \eq{Afb_flav} by,
\beq
\frac{c^{(8)}_A}{\Lambda^2}\lesssim \frac{0.2 }{\TeV^{2}}\, ,
\eeq
while for the case that there is no flavor invariance the best bounds on the parameter space are given by the previous analysis, since with the current data we can just exclude masses above 3 TeV in the parameter space.


\vskip.7cm

\section*{Acknowledgments}
The work of AP was partly supported by the projects FPA2011-25948, 2009SGR894 and ICREA Academia Program.
The work of JS was partly supported by the Spanish FPU Grant AP2006-03102, the project UNILHC PITN-GA-2009-237920 and the ERC Starting Grant Cosmo@LHC 204072.
The work of OD was partly supported by the FPU Grant  AP2007-00420.
OD thanks Roberto Franceschini and  Alvise Varagnolo for discussions.
\vskip1cm

\appendix

\section{Dimension six operators involving quarks}

Here we list the set of independent higher-dimensional operators involving SM quarks. 
As explained in section \ref{sec1}, we assume a flavor symmetry for the three  left-handed quarks $q_L$,  the three  right-handed down-quarks $d_R$,
and the two lightest right-handed up-quarks $u_R$, given by 
$U(3)_{q} \otimes U(3)_{d} \otimes U(2)_{u}$.
The top right-handed quark $t_R$ will be considered a singlet of the flavor symmetry.
We use the following notation.
We label with  $A$, $I$ and $F$   the  color, electroweak and  flavor index respectively in the adjoint representation. 
The contraction of the indices in the fundamental representation of these symmetries is understood within the fields in parenthesis, and  flavor indices can also be contracted with Yukawa matrices $Y_{u,d}$. 
We identify $T_A = \lambda_A/2$, being $\lambda_A$ the Gell-Mann matrices, and $\tau_{I} = \sigma_{I}/2$, where $\sigma_{I}$ are the Pauli matrices.

We  classify the operators  according to their expected suppression. 
First, we show the list of independent operators unsuppressed by Yukawa couplings (those generated in the massless quark limit). 
Following  the discussion of section \ref{sec1},
we separate these operators
as those of  first class   and second class, \eq{1st} and  \eq{2nd} respectively.
We finally   show the list of independent operators  suppressed by Yuwaka couplings.

      \subsection{First class operators}
    \subsubsection{Four-quark operators}
  \label{4fer}
   \begin{eqnarray}
 \label{O1uu}
\Op^{(1)}_{dd}&=&(\bar{d}_R\gamma^\mu d_R)(\bar{d}_R\gamma_\mu d_R)\nonumber\\
\Op^{(1)}_{ud}&=&(\bar{u}_R\gamma^\mu u_R)(\bar{d}_R\gamma_\mu d_R)
\quad\quad\quad\quad \quad\quad\quad 
 \Op^{(1)}_{td}=(\bar{t}_R\gamma^\mu t_R)(\bar{d}_R\gamma_\mu d_R)\nonumber\\
\Op^{(1)}_{uu}&=&(\bar{u}_R\gamma^\mu u_R)(\bar{u}_R\gamma_\mu u_R)
\quad\quad\quad\quad \quad\quad \quad
 \Op^{(1)}_{ut}=(\bar{u}_R\gamma^\mu u_R)(\bar{t}_R\gamma_\mu t_R) \nonumber\\
& \  &  \quad\quad\quad\quad \quad\quad\quad
\quad\quad\quad\quad \quad\quad\quad\quad
\Op^{(1)}_{tt}=(\bar{t}_R\gamma^\mu t_R)(\bar{t}_R\gamma_\mu t_R)\nonumber\\
\Op^{(1)}_{qu} &=& (\bar{q}_{L} \gamma^\mu q_{L}) (\bar{u}_{R} \gamma_{\mu}  u_R)
\quad\quad\quad\quad \quad\quad \quad
 \Op^{(1)}_{qt} = (\bar{q}_{L} \gamma^\mu q_{L}) (\bar{t}_{R} \gamma_{\mu}  t_R)  \nonumber\\
\Op^{(1)}_{qd} &=& (\bar{q}_{L} \gamma^\mu q_{L}) (\bar{d}_{R} \gamma_{\mu}  d_R) \nonumber\\
\Op^{(1)}_{qq}&=&(\bar{q}_L\gamma^\mu q_L)(\bar{q}_L\gamma_\mu q_L) 
\nonumber\\
\Op^{(3_W)}_{qq}&=&(\bar{q}_L\gamma^\mu \tau^I q_L)(\bar{q}_L\gamma_\mu \tau^I q_L)\nonumber \\
\Op^{(8_F)}_{qq}&=&(\bar{q}_L\gamma^\mu T^P q_L)(\bar{q}_L\gamma_\mu T^P q_L)\nonumber\\ 
\Op^{(8)}_{uu}&=&(\bar{u}_R\gamma^\mu T^A u_R)(\bar{u}_R\gamma_\mu T^A u_R)  \quad\quad\quad\quad
\Op^{(8)}_{ut}=(\bar{u}_R\gamma^\mu T^A u_R)(\bar{t}_R\gamma_\mu T^A t_R)
\nonumber\\
\Op^{(8)}_{dd}&=&(\bar{d}_R\gamma^\mu T^A d_R)(\bar{d}_R\gamma_\mu T^A d_R)\nonumber\\
\Op^{(8)}_{ud}&=&(\bar{u}_R\gamma^\mu T^A u_R)(\bar{d}_R\gamma_\mu T^A d_R)
\quad\quad\quad\quad
 \Op^{(8)}_{td}=(\bar{t}_R\gamma^\mu T^A t_R)(\bar{d}_R\gamma_\mu T^A d_R)\nonumber\\
\Op^{(8_C)}_{qq}&=&(\bar{q}_L\gamma^\mu T^A q_L)(\bar{q}_L\gamma_\mu T^A q_L)\nonumber\\
\Op^{(8)}_{qu} &=& (\bar{q}_{L} \gamma^\mu T^A q_{L}) (\bar{u}_{R} \gamma_{\mu}  T^A u_R)\quad\quad\quad\quad
 \Op^{(8)}_{qt} = (\bar{q}_{L} \gamma^\mu T^A q_{L}) (\bar{t}_{R} \gamma_{\mu}  T^A t_R)
 \nonumber \\
\Op^{(8)}_{qd} &=& (\bar{q}_{L} \gamma^\mu T^A q_{L}) (\bar{d}_{R} \gamma_{\mu}  T^A d_R)
\label{O8F}
 \end{eqnarray}
For physics involving only the first family quarks,  
that as  explained in  section~\ref{dijets} mainly corresponds to 
the  LHC dijet data $pp \rightarrow jj$, 
we can  reduce the above set of   four-quark operators
to   the set of \eq{4u}.
In this reduction, we have 
\bea
c^{(1)}_{uu} + \frac{1}{3} c^{(8)}_{uu} &\rightarrow&  c^{(1)}_{uu}\nonumber \\ 
c^{(1)}_{dd} + \frac{1}{3} c^{(8)}_{dd} &\rightarrow& c^{(1)}_{dd} \nonumber\\ 
c^{(1)}_{qq} - \frac{1}{12} c^{(3_W)}_{qq} + \frac{1}{3} c^{(8_F)}_{qq} &\rightarrow& c^{(1)}_{qq}\nonumber \\ 
c^{(8_C)}_{qq} + c^{(3_W)}_{qq}  &\rightarrow& c^{(8)}_{qq} 
\eea

  \subsubsection{Higgs-quark operators}
  \label{qv}
\begin{eqnarray}
 \Op_{Hu}  &=&  i(H^\dagger \overleftrightarrow{D}^\mu H) (\bar{u}_R \gamma_\mu u_R)\quad\quad\quad\quad\quad\quad 
  \Op_{Ht} = i(H^\dagger \overleftrightarrow{D}^\mu H) (\bar{t}_R \gamma_\mu t_R)  \nonumber\\
 \Op_{Hd} &=& i(H^\dagger \overleftrightarrow{D}^\mu H) (\bar{d}_R \gamma_\mu d_R)  \nonumber\\
 \Op^{(1)}_{Hq} &=& i(H^\dagger \overleftrightarrow{D}^\mu H) (\bar{q}_L \gamma_\mu q_L)
 \nonumber \\
 \Op^{(3)}_{Hq} &=& i(H^\dagger \tau^I \overleftrightarrow{D}^\mu H) (\bar{q}_L \gamma_\mu \tau^I q_L)
\eea
Notice that we just consider the antisymmetric part of the corresponding operators in \cite{Buchmuller:1985jz}. This is so because the symmetric part of the above operators can be put in the form $\partial_\mu(|H|^2)(\bar{\psi}\gamma^\mu\psi)$, where $\psi=(q_L,d_R,u_R,t_R)$. It can be shown that this kind of operators can be expressed in terms of operators appearing in  \ref{YukawaOp}.

    \subsection{Second class operators}
  \label{2ndclass}
   \begin{eqnarray}
\Op_{uB} &=&  (\bar{u}_R \gamma^{\mu} u_R)\partial^\nu B_{\mu \nu}
\quad\quad\quad\quad\quad\quad 
 \Op_{tB} =  (\bar{t}_R \gamma^{\mu} t_R)\partial^\nu B_{\mu \nu} \nonumber\\
\Op_{dB} &=&  (\bar{d}_R \gamma^{\mu} d_R)\partial^\nu B_{\mu \nu} \nonumber\\
\Op_{qB} &=&  (\bar{q}_L \gamma^{\mu} q_L) \partial^\nu B_{\mu \nu} \nonumber\\
\Op_{qW} &=&  (\bar{q}_L \tau^I \gamma^{\mu} q_L) D^{\nu} W^I_{\mu \nu}
 \end{eqnarray}
As explained in \cite{AguilarSaavedra:2008zc}  this kind of operators 
can be rewritten, by using the equations of motion (EOM) for the field strengths, 
 as  four-fermion operators involving quarks and leptons.
   Also notice that  operators involving gluons are not included since they can be rewritten as four-quark operators.
 
  \subsection{Yukawa-suppressed operators}
  \label{YukawaOp}
The Yukawa couplings break the flavor symmetry and generate extra dimension-six operators.
 Assigning to the $3\times 3$  Yukawa matrices, $Y_d$ and $Y_u$,
  the following quantum numbers under $G_F$: $Y_d\in {\bf ( 3,\bar 3,1)}$, 
 $\tilde Y_u\equiv(Y_u)_{i \tilde k}\in {\bf ( 3,1, 2)}$ $(i=1,2,3;\ \tilde k=1,2)$ 
and $\tilde Y_t\equiv(Y_u)_{i3}\in {\bf (3,1,1)}$, we can write 
the following $G_F$-invariant operators ($\widetilde{H} = i \sigma_2 H^*$):
  \begin{enumerate}[(i)]
  \item  
  \label{H3}
  \begin{eqnarray}
  \Op_{uH}&=&(H^\dagger H)(\bar{q}_L \tilde Y_u\widetilde{H} u_R)
  \quad\quad\quad\quad
   \Op_{tH}=(H^\dagger H)(\bar{q}_L \tilde Y_t\widetilde{H} t_R)\nonumber\\
  \Op_{dH}&=&(H^\dagger H)(\bar{q}_L Y_d H d_R)
 \end{eqnarray}
 \item 
 \begin{eqnarray}
 \Op_{uGH}&=&(\bar{q}_L\tilde Y_u\sigma^{\mu\nu}T^{A}u_R)\widetilde{H}G^A_{\mu\nu}
  \quad\quad\quad\quad
   \Op_{tGH}=(\bar{q}_L\tilde Y_t\sigma^{\mu\nu}T^{A}t_R)\widetilde{H}G^A_{\mu\nu}\nonumber\\
 \Op_{uWH}&=&(\bar{q}_L\tilde Y_u\sigma^{\mu\nu}\tau^{I}u_R)\widetilde{H}W^I_{\mu\nu}
  \quad\quad\quad\quad
  \Op_{tWH}=(\bar{q}_L\tilde Y_t\sigma^{\mu\nu}\tau^{I}t_R)\widetilde{H}W^I_{\mu\nu}\nonumber\\
 \Op_{uBH}&=&(\bar{q}_L\tilde Y_u\sigma^{\mu\nu}u_R)\widetilde{H}B_{\mu\nu}
  \quad\quad\quad\quad \quad
    \Op_{tBH}=(\bar{q}_L \tilde Y_t \sigma^{\mu\nu}t_R)\widetilde{H}B_{\mu\nu}\nonumber\\
 \Op_{dGH}&=&(\bar{q}_LY_d\sigma^{\mu\nu}T^{A}d_R)HG^A_{\mu\nu}\nonumber\\
 \Op_{dWH}&=&(\bar{q}_LY_d\sigma^{\mu\nu}\tau^{I}d_R)HW^I_{\mu\nu}\nonumber\\
\Op_{dBH}&=&(\bar{q}_LY_d\sigma^{\mu\nu}d_R)HB_{\mu\nu}
 \end{eqnarray}

\item

 \begin{eqnarray}
 \Op_{uDH}&=&(\bar{q}_L \tilde Y_u u_R)D_\mu D^\mu \widetilde{H}
  \quad\quad\quad\quad
   \Op_{tDH}=(\bar{q}_L \tilde Y_t t_R)D_\mu D^\mu \widetilde{H}\nonumber\\
 \Op_{dDH}&=&(\bar{q}_L Y_d d_R)D_\mu D^\mu H
 \end{eqnarray}
By applying the EOM of $H$ these operators could be rewritten as other operators of the list plus
four-fermion operators involving leptons.
  \item
\begin{eqnarray}
 \Op_{Hud}&=& i(\widetilde{H}^\dagger \overleftrightarrow{D}_\mu H)(\bar{u}_R \tilde Y^\dagger_u  Y_d \gamma^\mu d_R)
  \quad\quad\quad\quad
   \Op_{Htd}= i(\widetilde{H}^\dagger \overleftrightarrow{D}_\mu H)(\bar{t}_R \tilde Y^\dagger_t   Y_d\gamma^\mu d_R)
\end{eqnarray}
\item
   \begin{eqnarray}
\Op^{(1)}_{qud} &=& (\bar{q}_{L} \tilde Y_u u_{R}) (\bar{q}_{L} Y_d d_R)
 \quad\quad\quad\quad \quad\quad\quad 
 \Op^{(1)}_{qtd} = (\bar{q}_{L} \tilde Y_t  t_{R}) (\bar{q}_{L} Y_d d_R)\nonumber\\
\Op^{(8)}_{qud} &=& (\bar{q}_{L} \tilde Y_u T^A  u_{R}) (\bar{q}_{L} Y_d T^A  d_R)
 \quad\quad\quad\quad
  \Op^{(8)}_{qtd} = (\bar{q}_{L} \tilde Y_t T^A t_{R}) (\bar{q}_{L} Y_d T^A d_R)
 \end{eqnarray}
\end{enumerate}



\end{document}